\documentclass[12pt,preprint]{aastex}

\begin{document}


\title{The 2001 April Burst Activation of SGR 1900+14: X-ray afterglow emission}


\author{M. Feroci$^{1}$, S. Mereghetti$^{2}$, P. Woods$^{3,4}$,
C. Kouveliotou$^{3,4}$, E. Costa$^{1}$, D.D. Frederiks$^{5}$, 
  S.V. Golenetskii$^{5}$, K. Hurley$^{6}$,  
E. Mazets$^{5}$, P. Soffitta$^{1}$, M. Tavani$^{1}$}
\altaffiltext{1}{Istituto di Astrofisica
Spaziale e Fisica Cosmica - CNR, Rome, Italy}
\altaffiltext{2}{Istituto di Astrofisica Spaziale e Fisica Cosmica
- CNR, Sezione di Milano, Italy}
\altaffiltext{3}{NASA Marshall Space Flight Center, SD50, Huntsville,
  AL 35812}
\altaffiltext{4}{Universities Space Research Association}
\altaffiltext{5}{Ioffe Physico-Technical Institute, Russian Academy of
  Science, St. Petersbourg, 194021, Russia}
\altaffiltext{6}{Space Science Laboratory, University of
California, Berkeley, CA 94720-7450}

%\altaffiltext{1}{Visiting Astronomer, Cerro Tololo Inter-American Observatory.
%CTIO is operated by AURA, Inc.\ under contract to the National Science
%\altaffiltext{2}{Society of Fellows, Harvard University.}
%\altaffiltext{3}{present address: Center for Astrophysics,
%    60 Garden Street, Cambridge, MA 02138}
%\altaffiltext{4}{Visiting Programmer, Space Telescope Science Institute}
%\altaffiltext{5}{Patron, Alonso's Bar and Grill}


\begin{abstract}
After nearly two years of quiescence, the soft gamma-ray repeater
SGR 1900+14 again became burst-active on April 18 2001, when it
emitted a large flare, preceded by few weak and soft short bursts.
After having detected the X and gamma prompt emission of the flare,
BeppoSAX pointed its narrow field X-ray telescopes to the source in less than
8 hours.
In this paper we present an analysis of the data from this and
from a subsequent BeppoSAX observation, as well as from a set of RossiXTE
observations. Our data show the detection of an X-ray afterglow
from the source, most likely related to the large hard X-ray flare.
In fact, the persistent flux from the source, in 2-10 keV,
was initially found at a level $\sim$5 times
higher than the usual value. Assuming an underlying persistent (constant)
emission, the decay of the excess flux can be reasonably well described
by a t$^{-0.9}$ law. 
A temporal feature - a $\sim$half a day long bump - is observed 
in the decay light curve approximately one day after the burst onset. 
This feature is unprecedented in SGR afterglows. 
We discuss our results in the context of previous
observations of this source and derive implications for the physics of
these objects.
\end{abstract}

%\keywords{stars: neutron --- X-rays: stars}

\section{Introduction}

Soft gamma-ray repeaters (SGRs) are a small class of high energy
transients, comprising four confirmed members and a candidate one
(see e.g. Hurley 2001a for a recent observational review). These
sources usually manifest themselves through the emission of short
(few hundreds of milliseconds) bursts of hard-X/gamma-rays
recurring on variable timescales, from minutes to days to years.
All of the confirmed SGRs have been identified as steady X-ray
emitters (e.g., Hurley 2001a, and references therein) 
with luminosity in the range of
10$^{33}$-10$^{36}$ erg s$^{-1}$ (2-10 keV).
The X-ray radiation emitted by two of the SGRs, SGR 1806-20 and
SGR 1900+14, is coherently modulated with periods of $\sim$7.5 and
$\sim$5.2~s, respectively \cite{kouveliotou98,hurley99a}; both
sources show a secular spindown of $\sim$10$^{-10}$ s s$^{-1}$
(e.g., Mereghetti et al. 2000, Kouveliotou et al. 1998, 
Hurley et al. 1999a, Woods et al. 1999a). SGR
0526-66 exhibited coherent 8-s pulsations during the decaying part
of its famous 1979 March 5th giant flare \cite{mazets79}. These
pulsations were also seen during recent {\it Chandra} observations
of the source, albeit at a lesser statistical significance
\cite{kulkarni02}. Finally, no periodicity was found in the X-ray 
flux of SGR 1627-41 \cite{woods99b,hurley00b}.

One of the dramatic manifestation of the SGR sources are their giant flares.
Two sources have emitted such events so far: SGR 0526-66 (1979 March 5,
Mazets et al. 1979), and SGR 1900+14 (1998 August 27,
Hurley et al. 1999b, Feroci et al. 1999, Mazets et al. 1999a). 
These events are strikingly similar
(see, e.g., Mazets ey al. 1999a and Feroci et al. 2001a), very different from the
more frequent short recurrent bursts. In summary, both SGR giant flares
have light curves consisting of an initial short and very hard spike
followed by a much longer (several hundreds of seconds) tail, modulated
with the spin period of the neutron star.

The most important properties that distinguish these ``giant" flares from the
short recurrent bursts are their peak fluxes
($\sim$10$^{-3}$-10$^{-2}$ {\it vs}
10$^{-6}$-10$^{-5}$ ergs cm$^{-2}$ s$^{-1}$ in the short bursts),
their total energy
($>$10$^{44}$ ergs {\it vs} $\lesssim$10$^{41}$ ergs of the short bursts),
their duration ($\sim$300~s vs. few hundreds of milliseconds),
their periodically modulated time profile and the hard component in their
energy spectrum (in contrast to the thermal spectrum of the short bursts),
particularly in the very hard short spike emitted at the beginning of the
event.

Until recently, therefore, the distribution of SGR outbursts
appeared to be bimodal, comprising smaller outbursts and giant
flares. Recently, an event with 'intermediate' fluence and duration
of $\sim$0.5~s was detected
from SGR 1627-41 (Mazets et al. 1999b, see also Kouveliotou et al. 2001
for a general discussion on the distribution of properties of SGR
outbursts). In addition, on 2001 April 18, 07:55:12 UT, after
almost two years of burst-quiescence, SGR 1900+14 emitted a `brand new'
type of `intermediate' outburst that was detected by the BeppoSAX
Gamma Ray Burst Monitor \cite{guidorzi01}. The event had the
intermediate duration of $\sim$40~s, its light curve did not show
any initial hard spike, and was clearly spin-modulated. Also the
energetics appeared to be intermediate in the 40-700 keV energy
range, with a peak flux of $\sim$10$^{-5}$ erg cm$^{-2}$ s$^{-1}$
and a fluence of $\sim 1.5\times 10^{-4}$ erg cm$^{-2}$
\cite{guidorzi02}, corresponding (for isotropic emission at 10
kpc) to a peak luminosity of $\sim 1.3\times 10^{41}$ erg s$^{-1}$
and a total emitted energy of $\sim 1.9\times 10^{42}$ ergs.
Contrary to any large outburst from SGRs observed in the past,
this event was also observed at X-rays (2-26 keV) with the
BeppoSAX Wide Field Cameras, that also detected three weak short
bursts preceding the burst itself \cite{feroci01b}.
Unfortunately, the extreme brightness at X-rays activated the
self-protective automatic shutdown of the WFC instrument after
only 3~s, when a count rate of about 3$\times 10^{4}$ count
s$^{-1}$ was reached. The same event was detected by the
gamma-ray burst detector onboard {\it Ulysses} and by the
Konus-Wind experiment. The triangulation by the Third
Interplanetary Network provided an annulus consistent with the
position of SGR 1900+14 \cite{hurley01b}.

We present here observations carried out with the BeppoSAX
Narrow Field Instruments (NFI, Boella et al. 1997) and the Proportional
Counters Array (PCA, Jahoda et al. 1996) onboard the Rossi X-ray Timing Exporer
(RXTE) soon after the April 18 event. These observations started only
$\sim$8 hours after the flare, enabling us to detect an X-ray afterglow
fading into quiescence after a few days. We discuss in this paper
the temporal behaviour of the X-ray flux from this source. A companion paper
discusses the X-ray pulse properties and timing analyses \cite{woods03}
resulting from the same set of observations.

\section{Observations and Data Analysis}

\subsection{BeppoSAX NFI}

The detection and localization of the intermediate flare by the BeppoSAX Gamma
Ray Burst Monitor and Wide Field Camera unit \#1 prompted a follow-up
observation with the BeppoSAX NFI\footnote{Instruments of relevance here
are the LECS, Low Energy Concentrator Spectrometer effectively
operating in 0.1-4~keV
and the MECS, Medium Energy Concentrator Spectrometer, operating in
1.6-10~keV.},
as a part of our ongoing Target of Opportunity program for active
Soft Gamma-ray Repeaters.
Thanks to a remarkable effort of the BeppoSAX team,
the first observation started within less than 7.5 hours after the flare,
namely on 18 April 2001 at 15:10 UT, and ended on 19 April 2001, at 19:38 UT,
resulting in net exposure times of 46.2 and 20.4 ks for the MECS and
the LECS\footnote{For instrument safety reasons, the LECS was operated only
during the satellite night time.}, respectively.
A second pointing started on 29 April 2001 at 20:34 UT and ended
on 1 May 2001 at 08:25 UT, resulting in net exposure times
of 57.2 and 23.4 ks on the MECS and the LECS, respectively.

We analysed the data from the LECS and MECS instruments
starting from the cleaned event files provided by the BeppoSAX Science
Data Center.
The photons detected in the MECS during the first pointing (April 18)
were extracted in a circular area of 4$^{\prime}$ radius.
Given the low galactic latitude of this source,
we estimated the local background from an annulus
concentric with the source extraction region, of inner/outer radii
of 6$^{\prime}$ and 9$^{\prime}$, respectively.

\begin{figure}
%\centering
\rotatebox{-90}{
\plottwo{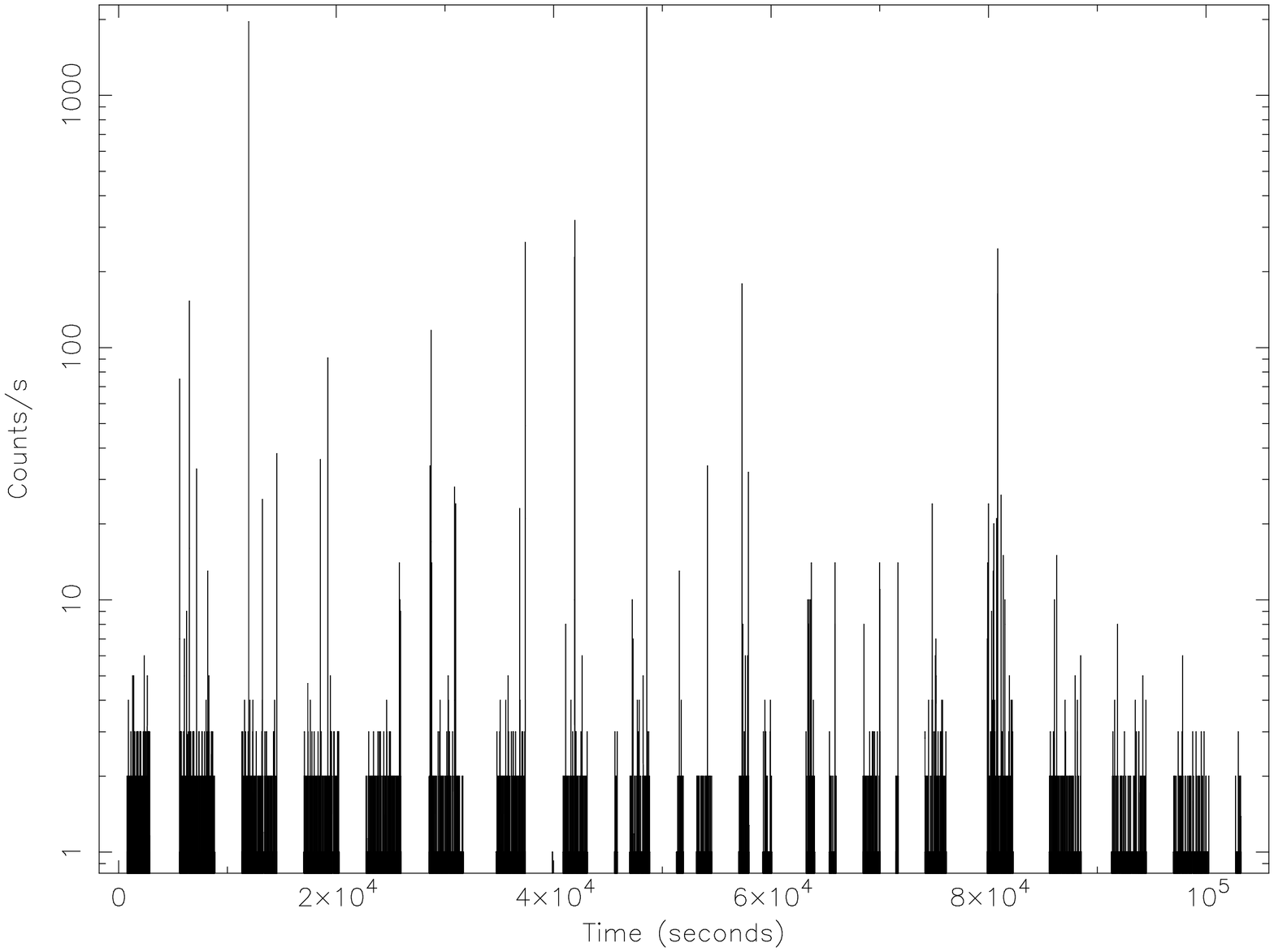}{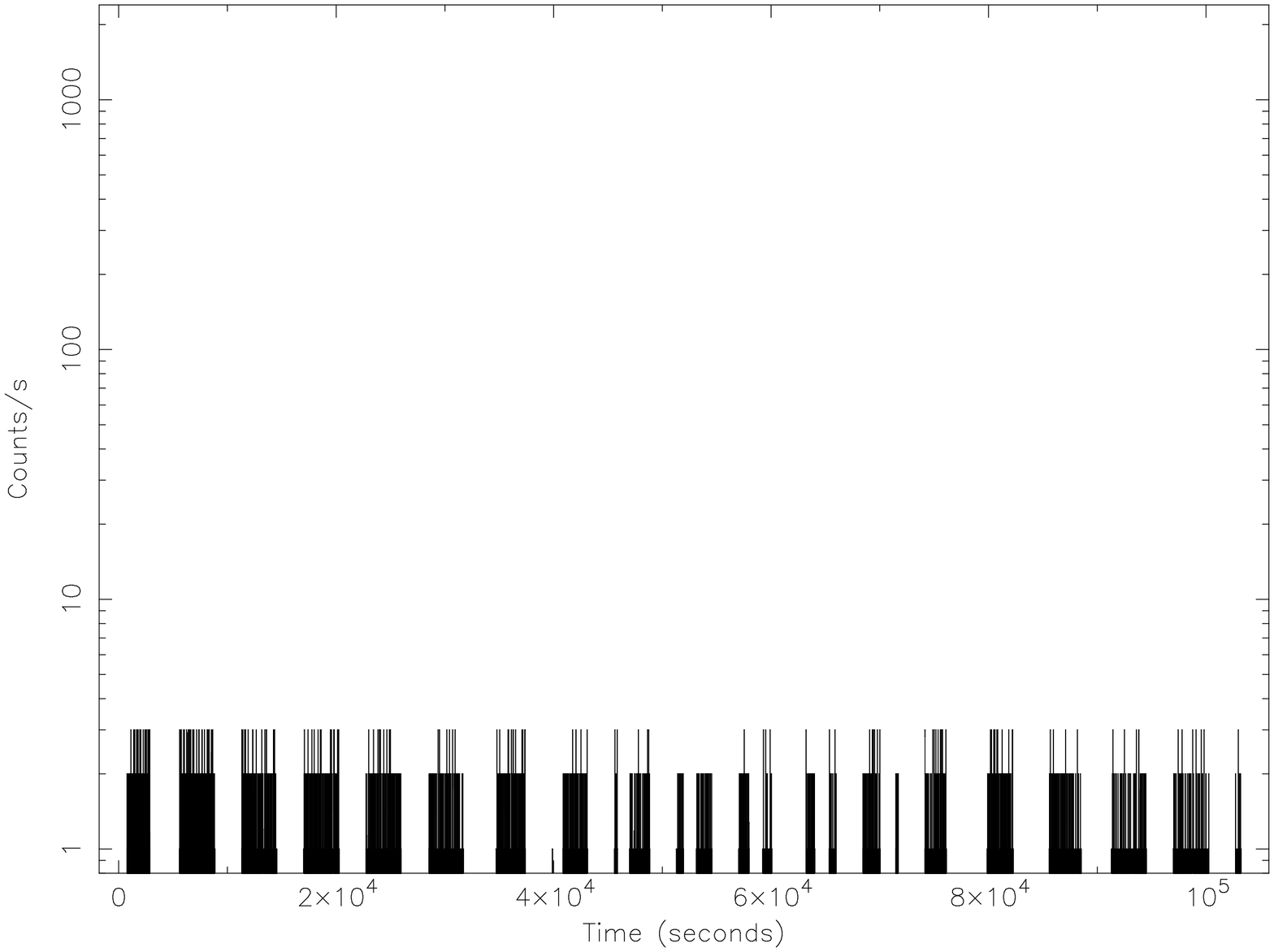}}
\caption{ 
Light curve of the April 18 BeppoSAX/MECS observation before
(left hand side) and after (right hand side) the removal of the 
bursts (see text for details). The bin size is 1 s.
\label{lc}}
\end{figure}

Further, we extracted a light curve of the MECS count rate, with a bin size
of 1~s, from which we clearly see a large number of bursts 
of different intensity and
duration from the source. To study the properties of the persistent 
emission of the source, we cleaned the event list from these bursts by
applying an intensity filter between 0 and 3 counts s$^{-1}$ (the
average source count rate is about 0.36 counts s$^{-1}$). This
`cleaning' procedure removed photons from about 70 bursts from
the photon list and left a net integration time of 46.1 ks. 
A plot of the MECS light curve before and after this procedure is
shown in Fig.~\ref{lc}.
The analysis of these bursts will be presented elsewhere.

\clearpage

\clearpage

The same procedures were applied to the LECS data using again extraction
regions of 4$^{\prime}$ radius and an annulus of 9$^{\prime}$ and 12$^{\prime}$
inner/outer radius, respectively. Due to the smaller exposure time, the
LECS missed several of the bursts observed with the MECS. 
Only 10 bursts were removed from its light curve, leaving a net
exposure time of 20.4 ks.

In order to convert the count rates into physical units, we
performed a simultaneous spectral fit combining the LECS (0.5-4.0
keV) and MECS (1.6-10.0 keV) data. Adopting a simple power law
with photoelectric absorption we find a satisfactory fit, with
reduced $\chi^{2}$=0.86 (91 degrees of freedom) and the following
spectral parameters: photon index $\Gamma$ = (2.57$\pm$0.08) and
N$_{H}$ = (4.41$\pm$0.25)$\times10^{22}$ cm$^{-2}$, resulting in
an unabsorbed flux of 3.4$\times10^{-11}$ erg cm$^{-2}$ s$^{-1}$
(2-10 keV). 
(Uncertainties on fit parameters are given at 90\% confidence level, 
whereas on data points they are 1-$\sigma$.)

\begin{figure}
\centering
\plotone{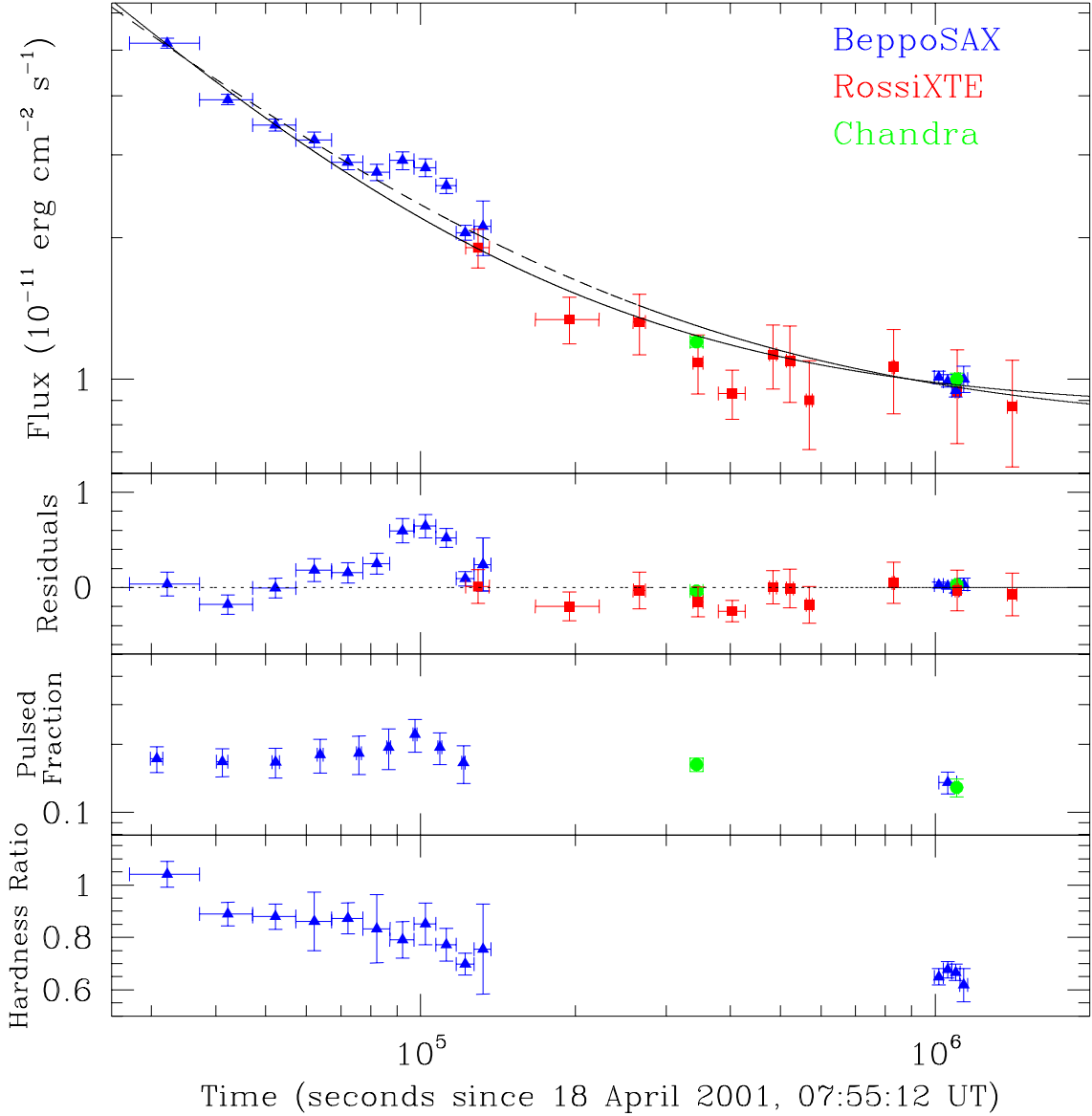}
\caption{
{\it Top Panel}: Temporal behaviour of the X-ray (2-10 keV) flux from
SGR 1900+14
in the aftermath of the April 18 2001 flare, as observed with the
BeppoSAX NFI, Chandra ACIS and the RossiXTE PCA,
together with the best-fit curves (see text for details) with a
constant plus a power law decay, including (dashed line) or
excluding (solid line) the data points in the bump.
{\it Second Panel}: Flux residuals from the fit performed excluding
the data points in the bump. The horizontal dotted line indicates the
zero. 
{\it Third Panel}: Light curve of the pulsed fraction during the two
BeppoSAX observations, as observed with the MECS, and during the two
Chandra ACIS observations (Woods et al. 2003).
{\it Bottom Panel:} Hardness ratio of the two BeppoSAX observations,
between the MECS energy ranges 4-10 keV and 1.6-4 keV.
(Uncertainties on all data points are 1-$\sigma$.)
\label{decay}}
\end{figure}

The burst-cleaned MECS photon list was then used to extract a light curve.
We choose time bins of 10$^{4}$~s to derive
the light curve that, converted to $cgs$ units using the counts to energy
conversion obtained from the above time-averaged spectrum,
is plotted in Fig.~\ref{decay} (full blue triangles, on the left hand side).
In order to check whether our counts-to-flux conversion procedure could be
significantly affected by any intra-observation spectral variability,
we computed a time-resolved hardness ratio between the MECS energy
ranges 4-10~keV and 1.6-4~keV (Fig.~\ref{decay}, bottom panel).
Indeed, we observe a general softening trend of the spectrum
across the observation but the spectral variation does not affect the
counts-to-flux conversion to more than a few percent, as also verified
through time-resolved spectra (see Mereghetti et al. 2003 for details).
Therefore, for simplicity we decided to use a single conversion factor for
the entire observation.

\clearpage

\clearpage

The same procedures were also applied to the data of the
second NFI observation (April 29). However, during this observation we
only detected a single burst,
thus the net exposure times of the cleaned data were 23 ks for the
LECS and 57 ks for the MECS.
The energy spectra from this observation cannot be fit with an absorbed
single power law. The resulting reduced $\chi^{2}$ of this
fit is 1.34 (72 degrees of freedom). Motivated by the general knowledge
on the energy spectra from these sources (e.g., Woods et al. 1999c), we then
added a blackbody component to the fitting function, that indeed improved
the fit: the reduced $\chi^{2}$ is now 1.06 (70 degrees
of freedom) with the following parameters: photon index
$\Gamma$ = 1.54$\pm{0.34}$,
N$_{H}$ = (1.9$\pm$0.3)$\times10^{22}$ cm$^{-2}$ and
kT = (0.60$\pm{0.05}$) keV.
The statistical confidence on the improvement in the fit obtained by
adding the blackbody component, as given by an F-test, is basically 100\%.
The unabsorbed 2-10 keV flux obtained using these spectral parameters is
0.98$\times10^{-11}$ erg cm$^{-2}$ s$^{-1}$.
A light curve was extracted for this second observation as well,
using 4$\times10^{4}$~s bin size,
due to the lower statistics. The four points obtained following this
procedure are plotted in Fig.~\ref{decay} (full blue triangles, right hand side).

We note that the large value of N$_{H}$ that we find in our first
observation - a factor of 2 larger than what usually found for this
source (N$_{H} \sim 2\times10^{22}$ cm$^{-2}$, e.g. Woods et
al. 1999c) - might depend on our choice of fitting with a simple power
law model. However, since this observation is the only one at such a high
flux level (among those carried out with N$_{H}$-sensitive instruments),
we cannot also eclude that this (transient) large N$_{H}$ might be intrinsic
to the source and in some way related to the flare.
A more extensive discussion of the spectral variations will be reported
in a forthcoming paper \cite{mereghetti03}.

\subsection{RossiXTE PCA}

Following the reactivation of SGR 1900+14, a sequence of RXTE
pointings was performed.  The RXTE observations began $\sim$34 hours
following the April 18 flare on 19 April 2001, 17:58 UT, and
continued intermittently over the next two weeks ending on 5 May 2001.
The RXTE observations were split into 13 separate pointing having a
total net exposure time with the PCA of 128 ksec.
Details on the RXTE observations are described in Woods et al. (2003).

The PCA data were cleaned by removing bursts and limiting the energy
range to 2-10 keV photons. The resulting data were binned (0.125 s
resolution) and the bin times were transformed to the solar system
barycenter using the FTOOL {\it faxbary}.  Each of the 13 segments were folded
on the spin ephemeris reported in Woods et al. (2003) and the {\it rms} pulsed
flux was measured for each of the resulting pulse profiles.

The RXTE/PCA is not an imaging instrument; its large field-of-view
(1$^{\circ}$ FWHM), therefore, includes flux not only from the SGR, but
also from the Galactic ridge (Valinia \& Marshall 1998) and other
X-ray transients (e.g.\ XTE J1906+09 [e.g.\ Wilson et al. 2002]).
Thus, under normal circumstances, it not possible to accurately
measure the flux level of the persistent emission from SGR 1900+14,
as the source contributes 0.62 counts s$^{-1}$ within the 2$-$10 keV band,
while the remaining cosmic flux constitutes {\it at least} 8 counts s$^{-1}$,
including all the above contributing factors.
Following Woods et al. (2001), we set out to
measure the phase averaged flux from the PCA observations by first
calculating the pulsed flux from the SGR which is not contaminated by
the dominant background.

Previously, it was shown that at most epochs the pulsed fraction
remained constant at $\sim$11\%.  Under the assumption of a
constant pulsed fraction at most times, Woods et al. (2001)
converted pulsed flux measurements to absolute source flux
by simply dividing the pulsed flux by the pulsed
fraction.  However, Woods et al. (2003) has shown that the pulsed
fraction following the April 18 flare rises significantly higher
to $\sim$18\% initially, then decays approximately linearly with
time approaching its nominal 11\% value.  To convert our PCA
pulsed fluxes to source flux values, we have fit the 2-10 keV
pulsed fraction measurements to a linear decay and interpolated
this model to estimate the pulsed fraction at the epochs of our
RXTE pointings.  These values were used to convert our pulsed flux
measurements into phase-averaged source flux values.

The flux data obtained following the procedure outlined above were then
input into the same plot as the BeppoSAX data (Fig.~\ref{decay}, full red squares).
The first RXTE data point partially overlaps the
last BeppoSAX point, and it is consistent with it, providing a
confirmation of the goodness of the above procedure.
The RXTE data points nicely fill the gap between
the two BeppoSAX observations and allow us to monitor the return of the
source flux to quiescence.

\section{The flux history}

The data points for the flux history of SGR 1900+14 after the April 18
flare summarized in Fig.~\ref{decay} also include the flux
measurements obtained by
the instrument ACIS-S3 on the Chandra X-ray Observatory on April 22 and 30
\cite{kouveliotou01}. The overall set of observations provide a concise history
of the continuous flux decay of the source. Furthermore, they confirm the
cross-calibration between the different instruments, as the
overlapping data points indicate.

We have fitted the observed decay trend with a function composed
by a constant, accounting for the steady X-ray emission, plus a
power law decay - $\sim t^{-\alpha}$ - accounting
for the excess emission. As can be seen from the figure, this function
describes the data rather well, with the following fit parameters:
${\alpha}$=0.89$\pm$0.06,
constant=(0.78$\pm$0.05)$\times 10^{-11}$ erg cm$^{-2}$ s$^{-1}$.

We note that in a previous communication by us
\cite{feroci01c,feroci02} we indicated a decay index of about
0.6. However, in that case we only used a limited data set
and did not include the underlying constant flux
component in the fit. That index is still valid if the fit is
limited to  $t\lesssim$ 3$\times$10${^5}$ s and the constant flux
is not considered.

Based on the above fitting function, we can derive the total
(unabsorbed) energy emitted during the X-ray afterglow, assuming
isotropy. To this purpose we consider the time segment from the
end of the burst, $t\simeq$40~s in our reference frame, up to the
time when the afterglow extinguishes itself, that is when the
total flux returns to its usual value (approximately  10$^{-11}$
erg cm$^{-2}$ s$^{-1}$). We estimate this time to be at
$t\simeq$10$^{6}$~s. Integrating the best fitting function over
this time interval we get $\sim$2.1$\times 10^{-5}$ erg cm$^{-2}$,
in the energy range 2-10~keV, of which about 0.8$\times 10^{-5}$
erg cm$^{-2}$ is attributable to the (underlying) persistent
source emission (as estimated from our fit), bringing the net
afterglow energy output to $\sim$1.3$\times 10^{-5}$ erg cm$^{-2}$,
corresponding to approximately 1.6$\times 10^{43}$ ergs at 10
kpc. \footnote{It should be noted that different models may be used
to fit the decay data. In particular, Lenters et al. (2003)
fit a power-law to the total (afterglow + persistent emission) 
net source flux, then {\it later} 
subtracted the nominal persistent flux level to estimate the energy
output in the afterglow. They obtain an afterglow energy
approximately a factor of 2 smaller, in the same energy range. 
In our estimation, roughly
half of the energy output is inferred from back-extrapolation of
the best fit model for epochs between the burst and the start of
the BeppoSAX observation.} This number may be compared to the
energy output during the burst itself, available to us in the
40-700~keV range, 1.5$\times 10^{-4}$ erg cm$^{-2}$, showing that
it is approximately  a fraction of 10\% of it.

\subsection{The bump at $t\sim10^{5}$~s}

Although the simple power law function globally fits the data
rather nicely, the reduced $\chi^{2}$ of the fit is in excess of
3, and the derived constant flux emission underestimates the value
previously reported for the quiescent status of this source
(e.g., Hurley et al. 1999a). 
As it is evident from Fig.~\ref{decay}, one of the sources
of the high value for the $\chi^{2}$ is the bump occurring in the
light curve during the first BeppoSAX pointing at $t\sim10^{5}$~s.
We have studied this flux excess in more detail, searching for a
possible instrumental origin, a dependence on the source
extraction region, or the light curve bin size, and found none. We
also checked the data from the two MECS units independently, and
find that it appears in both. The possibility that it may be
related to an increased flux of solar protons (actually detected
by the Konus detectors), due to the ongoing intense solar
activity, is excluded by the study of the background
over the entire detector. Our conclusion, therefore, is that the
bump in the light curve is real.

Having excluded an instrumental origin, we have investigated the
nature of the bump, as being due to the SGR source itself. First,
we have studied the spectral properties. Already from the
time-dependent hardness ratio of the MECS data (Fig.~\ref{decay}),
it appears to have no particular spectral signature. A more
detailed spectral study (see Mereghetti et al. 2003 for details)
confirms that indeed no significant variation in the spectrum is
detected at the time of the bump. We have then checked the
possibility that the flux increase could be due to an X-ray tail
of an undetected (due to Earth occultation) high fluence, short
burst. This was done in two ways. First, using the data from the
Ulysses and Konus Gamma Ray Burst detectors, that do not suffer
from the Earth occultation problem. Unfortunately, at the time of
the observation the Sun was particularly active, strongly
decreasing the sensitivity of those instruments. However, Konus
detected few events from this source with fluences as low as
5$\times10^{-7}$ erg cm$^{-2}$ just about 3 hours before the bump,
but nothing is reported between that time and the time of the
bump. The upper limit on the fluence of a short ($\sim$0.5 s)
burst provided by Ulysses in 25-150 keV is approximately 10$^{-6}$
erg cm$^{-2}$.

Another way is to look at the pulsed fraction. In fact, Lenters et
al. (2003) have noted that the pulsed fraction largely increases
during two extended X-ray tails of the high fluence bursts, up to
a factor of 2. We have therefore studied the temporal behavior of
the pulsed flux during this observation (see Woods et al. 2003 for
details) and found that it follows a similar time behavior as the
non-pulsed flux. However, when one derives the time history of the
pulsed fraction (third panel of Fig.~\ref{decay}) finds that it
shows a bump as well. Unfortunately, the large statistical
uncertainties prevent us to derive any firm conclusion about it,
but, taken at face value, the data on the pulsed fraction during
the first BeppoSAX observation suggest only a small increase, if
any, in the pulsed fraction at the same time of the bump in the
total flux.

Assuming, following Lenters et al. (2003), that the
burst/afterglow fluence ratio is approximately constant, the
excess energy allows us to roughly estimate the fluence of a putative
undetected burst, whose afterglow may have been responsible for
the bump. To this purpose, we derived an estimate of the excess energy
emitted by SGR1900+14 during the bump. We added to the analytical
description used above - a power law plus a constant - a gaussian
line, with the only purpose of fitting the data
phenomenologically. The fit is indeed statistically very
satisfactory, giving a reduced $\chi^{2}$ of 0.8 (22 d.o.f.),
therefore we assume it is appropriate to derive the properties of
the bump. The $\sigma$ of the gaussian turns out to be
approximately 1.7$\times10^{4}$~s, the peak is about
0.76$\times10^{-11}$ erg cm$^{-2}$ s$^{-1}$ and the energy fluence
associated with it 3.3$\times10^{-7}$ erg cm$^{-2}$. The energy in
the bump is therefore only a few percent of the total energy emitted
by the source in the afterglow. Using this value for the fluence
in the afterglow, the fluence of an undetected burst should have
been of the order of several times $10^{-6}$ erg cm$^{-2}$, in
contrast with Konus and Ulysses data. We note, also, that if as
in the other known cases, the effect of a burst was a power law-like
X-ray tail, then the shape of the bump - well described with a
gaussian - tends to exclude such an origin. In fact, by
superposing two power law tails it is not possible to obtain the
slow rise that we observe in the bump, and a decay index greater
than 2.5 would be needed for the second power law in order to
match the fast decay of the bump, and the fit would not be
satisfactory even in this case. In addition, the increase in the
pulsed fraction that we possibly observe is quantitatively much
smaller than what reported by Lenters et al. (2003) from two other
observations. Based on these considerations, we assume as our
working hypothesis, that the bump was {\it not} due to an
undetected burst.

Having identified the bump as an `anomaly' in the flux decay, we
now attempt to derive a decay law that is not affected by the
bump. We fit the decay data excluding those points that appear to
constitute the bump itself (by visual inspection, from the 6th to
the 9th BeppoSAX data point). The resulting best-fit curve is the solid line
overplotted to Fig.~\ref{decay}. In this case the reduced
$\chi^{2}$ is rather satisfactory - 1.09 (21 d.o.f.) - and the fit
parameters change to the following values: $\alpha$=1.02$\pm$0.07,
constant=(0.86$\pm$0.05)$\times10^{-11}$ erg cm$^{-2}$ s$^{-1}$.
These parameters, although not dramatically different from the
previous ones, allow the fitting curve to better follow the decay
given by the RXTE and Chandra points at times between 2$\times
10^{5}$ s and 10$^{6}$~s and give a higher quiescent flux, more in
accord with previous observations.
From the fit residuals (second panel of Fig.~\ref{decay}) it appears
that all the data points between $t\sim 2\times10^{5}$~s and 
$t\sim 6\times10^{5}$~s are systematically overestimated from the 
analytical law (although none of them by more than 2$\sigma$).
This might well be a symptom that the (empirical) analytical law that
we used to describe the flux decay may not be completely adequate.

It is also interesting to note that the power law normalization
value that we derive for t=0 (the time of the flare) indicates
a 2-10 keV flux of the order of 5$\times 10^{-7}$ erg cm$^{-2}$
s$^{-1}$.
The average flux detected by the BeppoSAX Wide Field Camera in the
first seconds of the prompt event, in the same energy band, was just
about of the same order.
This means that the backward extrapolation of the flux of the
X-ray afterglow roughly matches
the intensity of X-ray emission during the burst, at least over the
first few seconds (although it is reasonable to guess that the prompt
X-ray flux may have reached a brighter peak at later times).
In order to verify whether such a back extrapolation could be
supported or denied by other observations, we checked the
available data from the All Sky Monitor onboard RXTE
(ASM, Levine et al. 1996), that
monitors the source several times each day, thus possibly filling
the observational gap between the flare and the start
of the observations with BeppoSAX. We found that the earliest
observation after the time of the flare is a single dwell (90~s) at
$ t \sim 1.5 \times 10^{4} $~s, and the ASM instrument did
not detect the source. A 2-$\sigma$ upper limit is about
$7.5 \times 10^{-10}$ erg cm$^{-2}$ s$^{-1}$ (Al Levine, 2002,
Priv. Comm.), whereas the flux at that time
implied by the back extrapolation of both our decay laws is at a level of
about $10^{-10}$ erg cm$^{-2}$ s$^{-1}$, thus compatible
with the ASM non-detection.

\section{Discussion}

Our detection of an X-ray afterglow after an SGR flare is not
unique. Three other cases were reported in the literature so far,
all of them from this same source: after the 1998 August 27 giant
flare \cite{woods01}, and after two `unusual' short bursts on
1998 August 29 \cite{ibrahim01} and 2001 April 28
\cite{lenters03}. The four cases have several similarities that
may lead to the idea that they are indeed different manifestations
of the same phenomenon. In the following, we therefore discuss to
what extent they are actually similar.

The most basic property that all these four bursts share is a large fluence:
their fluence is in excess of 10$^{-6}$ erg cm$^{-2}$
(e.g., Lenters et al. 2003),
whereas the common short bursts from this source usually have fluences
between 10$^{-10}$ and 10$^{-7}$ erg cm$^{-2}$ (e.g.,
Gogus et al. 1999). In addition, to our knowledge, they also share the
observational bias of being all of the large-fluence flares that have
been followed by an observation with sensitive X-ray instruments
shortly (i.e., hours) after the
burst. So, with the limits given by this very small sample of events,
there may be an indication that a soft X-ray afterglow follows {\it
  every} large-fluence burst, and possibly even {\it every burst}. If the
burst-afterglow energy ratio is roughly constant, as suggested by
Lenters et al. (2003), then this would simply result in the
undetectability of the afterglow of `standard bursts' with the
instrumentation used so far.

In all the four cases the shape of the afterglow decay
can be reasonably well represented by a power law
$I(t) \propto t^{-\alpha}$, and is accompanied
by a spectral softening (possibly due to a varying relative intensity
of the blackbody and the power law spectral components, see Mereghetti et
al. (2003) for a more detailed discussion on this issue in the case of
the April 18 event).
The index of the decay is found in a relatively narrow range in all
the four events: the two short ones had the slowest decay - 0.6 -
\cite{ibrahim01,lenters03}\footnote{
Actually, Lenters et al. (2003) described the tail of the April 28 event
with a power law folded with an exponential cut
off. However, considering an energy range 2-10 keV, the value for the
index should be reasonably well approximated by 0.6.},
while for the August 27 event Woods et al. (2001) found an index of
0.7 (but without including the persistent emission in their fitting
function, therefore that index should be regarded as a lower limit,
in absolute value, when compared to our results).
The difference in decay index between the two short and
smaller-fluence bursts ($\sim$0.6 for August 29 and April 28) and the
two long and large-fluence ones ($>$0.7 for August 27 and $\sim$0.9
for April 18) is therefore significant, but at this time the very small
number of events prevents us to draw any conclusions about possible
correlations of the decay index with the fluence.

A peculiarity that certainly distinguishes the afterglow of the
April 18 event from the others is the detection of a bump in the
light curve. The tails of the two short bursts (August 29 and
April 28) were indeed much shorter ($\sim$10$^{3}$~s). They show some
deviation from a simple power law (i.e., an exponential cut-off,
Lenters et al. 2003), but nothing resembling the bump feature seen
following the April 18 event. In the case of
August 27, the afterglow was detected over a much longer timescale
($\sim$40 days, Woods et al. 2001) with no evidence for a bump on the
scale of $\sim$half a day, although the sparse observations early in
this decay cannot rule it out.

What caused the bump? In the {\it magnetar} model (e.g., see Lyubarsky,
Eichler and Thompson 2002, and references therein) the long
duration (e.g., of the order of few 10$^{6}$ s) afterglow emission
is interpreted in terms of thermal emission from a localized
region on the neutron star surface. This is due to an \emph{in
situ} heating of the crust by a decaying magnetic field.
The region
on the surface is determined by the location of the crust crack
that originated the flare itself. This model can reproduce the
properties of the afterglow after the August 27 event, although
the simple picture may already require some complications when
data about the tails of the two short bursts are considered (see
Lenters et al. 2003 for details). Our detection of a bump in the
light curve possibly requires more of them.\footnote{Indeed, in
the magnetar model there are conditions related to sharp
variations in the subsurface heat conductivity in the the neutron
star, from which one can expect deviations from a pure power law
decay, even in the apparent form of a bump. Alternatively, the
possibility of gamma-ray quiet releases of energy is not ruled out
by the model. (D. Eichler \& Y. Lyubarski, Priv. Comm., and
Thompson \& Duncan, 1996)}

As we have shown, the possibility that the (small) additional energy
observed in the bump was injected by an unseen burst is unlikely,
unless it was ``star-occulted''.
In addition, the time behavior of the pulsed fraction requires a
geometrically extended phenomenon. In fact, although the energy
involved by the bump is a small fraction of the total afterglow
energy, its peak flux is
approximately 30\% of the `bump-subtracted' afterglow. If such an
increase in the total flux was entirely due to enhanced emission
only at the polar cap, the pulsed fraction would have increased by
about a factor of 2. This is not compatible with our data. On the
other hand, a bump excess emission due only to non-pulsed flux
would have resulted in a decrease of the pulsed fraction by
$\sim$25\%, contrasted by our data as well. We are therefore left
with the conclusion that the bump was due to a (possibly slightly
different) increase in {\it both} components. This is maybe easier
to obtain if the emission region is extended, than with a
fine-tuned location of a small emitting area on the star's
surface.
Finally, the purely non-thermal spectrum of the early afterglow in
this scenario
requires, as in the case of the August 27 event, a reprocessing of the
photons through an energetic pairs atmosphere.

An alternative scenario may be suggested by noting the similarity
of this phenomenon with the X-ray afterglow observed from the
gamma-ray bursters (GRBs). The X-ray afterglow in that context is
interpreted in terms of (mostly) synchrotron emission by an
electron population accelerated to relativistic energies by a
propagating shock front (e.g., Dermer \& Chiang 1998, and references
therein). In that case, the
observation of bumps in the otherwise smooth power law decay of
the X-ray flux is usually interpreted as due to a clumpy
interstellar medium, `refreshing' the shock when it encounters a
density increase (e.g., Lazzati et al. 2002). 
A similar scenario, likely applicable also in a
magnetar context, may be adapted to our case as well. The
advantage of it is that a non-thermal energy spectrum
is naturally explained by the electron acceleration and emission
mechanisms, and an extended emission region (required by our
observation of a weak or null change in the pulsed fraction at
least across the bump) may easily be the result of a non- or
mildly-collimated particle ejection and a small Lorentz factor of
the electron population, thus limiting also the Doppler boosting. We
remind, also, that a similar scenario was already thought to be at
work to explain the hard spike of the August 27 event, and the
transient radio emission observed a week later \cite{frail99}.
Actually, a major difference between that flare and the event
discussed here is the presence of the very hard and short spike at
the beginning of the flare, possibly the radiative signature of the
particle ejection \cite{feroci01a}. If this was indeed a signature of
the particle acceleration, then we need to assume that here the
first spike was not seen because it was Doppler-boosted in a direction
away from the observer,\footnote{It should be noted, however, that
for the August 29 event this hypothesis is directly contrasted by
the phase of the burst, coincident with the pulse maximum. This
problem does not apply to the April 18 and 28 events, occurring at
pulse minima \cite{lenters03,woods03}.} and we see the afterglow
only when the beaming is not effective anymore, due to a decreased
Lorentz factor. Again on the GRB-like behaviour, it is interesting
to note that we have shown that our data are compatible with an 
hypothesis of
continuity between the X-ray emission during the flare and in the
afterglow, just as it often happens in GRBs. In that context, the
afterglow-flare fluence ratio ranges from few percents up to
almost 30\% \cite{frontera00}, conistent with the $\sim$10\%
that we find for the April 18 event from SGR 1900+14.

In conclusion, a power law -like decaying X-ray afterglow seems to
be a `standard' consequence of large-fluence flares from SGR
1900+14, although only four cases have been observed. However, the 
details of such a phenomenon appear to change significantly from 
one case to another. Here we presented an X-ray afterglow decaying faster than
the previously reported events, and showing a bump in the light
curve, raising new questions to be addressed by the interpretative
models. 
The continued or future operation of missions like HETE-2, RossiXTE, 
INTEGRAL and AGILE, combined with the launch of Swift at the end 
of this year, with its capability of rapid X-ray follow-ups, 
will likely improve the chance of detecting new SGR afterglows and 
study their detailed behaviour. Of course, only in the case of 
cooperation by the SGR sources.

\acknowledgments The authors warmly thank Jean in 't Zand (who
promptly identified the flare in the Wide Field Cameras as coming
from SGR 1900+14), the BeppoSAX Mission Scientist and Mission
Planners team, the Operation Center and the Mission Director for
their remarkable efforts that made it possible to carry out our
observation so shortly after the flare. Without their
contribution, most likely the X-ray afterglow would not have been
detected. We also thanks Chris Thompson for useful comments on the
magnetar interpretation. KH is grateful for Ulysses support under
JPL Contract 958056, and for Ulysses-BeppoSAX collaborative work
under NASA Grants NAG5-9126 and NAG5-10710.




\end{document}